\documentclass{article}

\usepackage{arxiv}
\usepackage[utf8]{inputenc} 
\usepackage[T1]{fontenc}    
\usepackage[colorlinks=true, linkcolor=blue, citecolor=blue, urlcolor=blue]{hyperref}       
\usepackage{lipsum}
\usepackage{url}            
\usepackage{booktabs}       
\usepackage{amsfonts}       
\usepackage{nicefrac}       
\usepackage{microtype}      
\usepackage{graphicx}
\usepackage{amsmath}
\usepackage{physics}
\usepackage{booktabs}
\usepackage[capposition=top]{floatrow}
\usepackage{natbib}
\usepackage{doi}
\usepackage{caption}
\title{Determination of commensurabilities in the dynamics of the Pluto-Charon‘s satellites}

\date{March 3, 2024}

\author{
  Baptiste Desoubrie \\
  IMCCE, Observatoire de Paris, PSL Research University, CNRS, Sorbonne Université, Université de Lille \\
  77 avenue Denfert-Rochereau, 75014, Paris, France \\
  \texttt{baptiste.desoubrie@obspm.fr} \\
  \And
  Alain Vienne \\
  IMCCE, Observatoire de Paris, PSL Research University, CNRS, Sorbonne Université, Université de Lille \\
  77 avenue Denfert-Rochereau, 75014, Paris, France \\
  \texttt{alain.vienne@univ-lille.fr}
}

\renewcommand{\shorttitle}{Determination of commensurabilities in the dynamics of the Pluto-Charon‘s satellites}

\begin{document}
\maketitle

\begin{abstract}
The Pluto-Charon system is a binary system, characterised by its high mass ratio and its tidally locked synchronous orbit. Surrounding the central binary, four satellites - Styx, Nix, Kerberos and Hydra - follow near-circular and near equatorial orbits. Mutual tidal forces have probably a great influence on their long term dynamics. Several studies have pointed out some commensurabilities, sometimes using the fast Fourier transform technique. We intend to precisely identify the commensurabilities between the Pluto-Charon satellites, using the tool of fine frequency analysis, not only to investigate known commensurabilities but also to explore new ones. We used a combination of analytical and numerical approaches. The epicyclic theory of Lee and Peale served as an initial description of the motions and a comparative benchmark for numerical integrations. These integrations were performed using the IAS15 integrator within the N-body orbital integrator, REBOUND, considering both 3-body with the central binary and a single moon, and complete 6-body models. Additionally, we applied Frequency Map Analysis to determine precisely the numerical values of the system's fundamental frequencies. Our fine frequency analysis successfully identified all fundamental frequencies of the system, along with a detailed examination of the associated uncertainties. The findings, obtained with both 3-body and 6-body numerical models, align well with predictions from the epicyclic analytical theory. Notably, we highlight commensurabilities previously reported and introduce novel ones. Furthermore, we demonstrate how smooth alterations of the initial conditions can lead to the observation of libration in specific arguments, providing insights into the dynamical behavior of such a complex system.
\end{abstract}


\section{Introduction}
The Pluto-Charon's system is very interesting from a dynamical point of view. Firstly, it looks like a binary system, with a mass ratio about one-eighth between the two main bodies Pluto and Charon. Four small satellites follow near-circular, near-equatorial orbits around that inner binary: Styx, Nix, Kerberos and Hydra. It is rather common for a binary star system but unusual in our solar system. Nix and Hydra were discovered in 2005 by \citet{Weaver2006}. The smaller ones, Kerberos and Styx, were discovered later \citep{Showalter2011, Showalter2012}.
Secondly, Pluto and Charon are tidally locked: they move synchronously on a nearly circular orbit with a period of $ P_\mathrm{bin} \approx 6.3876$ days \citep{Buie2012}. Undoubtedly, tidal forces have a great influence in the long term dynamics of the four small satellites.
Thirdly, the system approaches a mean motion commensurability ratio of 1 : 3 : 4 : 5 : 6, but does not closely align with it. More generally, close commensurabilities could affect the system's dynamics if near enough. Identifying them is essential for a thorough understanding of the system's dynamics. Table \ref{table:orbital_parameters} summarizes the main characteristics of the Pluto-Charon system bodies.
\\
\\
\begin{table}[ht]
\centering
\caption{Main parameters of the Pluto-Charon system objects}
\begin{tabular}{lccc}
\hline \addlinespace[1pt]
Body & GM (km$^3$ s$^{-2}$) & $a$ (km) & Period ($P_\mathrm{bin}$)\\ \addlinespace[1pt]
\hline \addlinespace[2pt]
Pluto & \phantom{$\leq$ }869.6\phantom{00\,0}     & 19\,596 & 1\phantom{.000\,0}\\
Charon & \phantom{$\leq$ }105.9\phantom{00\,0}    & 17\,470 & 1\phantom{.000\,0}\\ 
Styx & $\leq$ \phantom{00}0.001\,0                & 42\,656 & 3.156\,6 \\
Nix & \phantom{$\leq$ 00}0.003\,0                 & 48\,694 & 3.891\,3 \\
Kerberos & \phantom{$\leq$ 00}0.001\,1            & 57\,783 & 5.036\,3 \\
Hydra & \phantom{$\leq$ 00}0.003\,2               & 64\,738 & 5.981\,0 \\
\hline
\end{tabular}
\floatfoot{Semi major axes $a$ are relative to the system's barycenter. All the parameters are adopted from \citet{Brozovic2015}.}
\label{table:orbital_parameters}
\end{table}
\noindent
The formation of the system remains uncertain. A giant impact between two transneptunian objects that would have led to the formation of the Pluto-Charon binary is the most plausible scenario \citep{Canup2021}. This event may have provided enough material to make the small moons within a circumbinary disk of debris \citep{Stern2006, Kenyon2014, Bromley2020, Kenyon2021}. Due to the tidal interactions, the binary expands in a few million years until Pluto and Charon are tidally locked, to the present separation $a_{\mathrm{bin}} \approx 19600 \text{km}$. As with the expansion of the binary, the regions of mean motion resonances (MMR) (1 : n) moves outward \citep{Farinella1979}.

Within the system dynamics, the primary perturbations arise from the non-axisymmetric potential of the central binary and mutual interactions among the satellites. While several other effects might impact the system's dynamics, their significance within our study's timescales remains marginal. Indeed, radiation pressure is negligible given the high densities of the satellites. Likewise, even though the satellites irregular shapes could introduce perturbations, the mutual distances and masses suggest these would be minor. The system's compactness, with satellites situated well within the Hill's sphere ($\sim 0.007$ to $0.011 \, R_\text{Hill}$), emphasizes the primary role the non-axisymmetric potential plays in the system's dynamics. External gravitational influences from the Sun and other planets might lead to secular motion, but these would manifest over timescales far exceeding our study's scope. Even then, such perturbations would have no substantial effect \citep{Michaely2017}. A broader overview of the different timescale effects is detailed in Fig.~2 of the same reference.
\\
\\
\noindent
Limiting the dynamics to the non-axisymmetric potential of the central binary, \citet{Lee2006} have modelled the orbit of a massless particle around a zero-eccentricity binary system. Their solution is the superposition of a circular motion around the center of mass, an epicyclic motion, forced oscillations due to the non-axisymmetric components of the binary's potential, and a vertical motion. We will use this representation for comparatives purposes.
\\
\\
\noindent
\citet{Woo2020} developed a Fast Fourier Transform (FFT) technique to derive the frequencies identified by the epicyclic model. They found an important deviation in the precession of Styx, which they attributed to the effect of the 3:1 mean-motion resonance with the binary. \citet{Gakis2022} use numerical integrations to show that the time-dependent, non-axisymmetric potential of the Pluto-Charon binary induces irregular patterns to the orbits, which may explain the observed differences between sets of orbital elements obtained by different previous studies. Furthermore, they have used the FFT technique for both the outcomes of the semi-analytic model and their n-body simulation \citep{Gakis2023}. They have found the main frequencies predicted by the semi-analytical model and have shown that mutual interactions have a significant effect with long period terms. They have also identified peaks in the low-frequency regime, due to the interactions between the small moons.
\\
\\
\noindent
In these previous works, the FFT technique was precise enough to detect and separate the main frequencies and some of their harmonics. In the present paper, we want to go further and identify the main behaviours of the mutual interactions, such as possible resonances. Studying the commensurabilities in a dynamical system is the first step towards finding the way to explore it. Thus, our aim is to investigate some known commensurabilities, but also to find new candidates. We use the tool of a fine frequency analysis. 
\\
We first recall in Sect.~\ref{section2} the tools we use. We start with the epicyclic theory of \citet{Lee2006}, for which we recall the formulae. They are useful to understand the outcomes of the numerical integrations we introduce just after. The fine frequency analysis is then presented, with a deep discussion of the uncertainties. The fundamental frequencies are identified and presented in Sect.~\ref{freq_section}. We are then able to present some known commensurabilities and explore new ones in Sect.~\ref{section4}. Finally, the last section gives conclusions and some perspectives.

%
\section{Analytical and numerical tools}
\label{section2}

\subsection{The epicyclic theory} \label{epicyclic_section}
\citet{Lee2006} have developed an analytic theory that describe the trajectory of a particle around a central binary, in order to take into account the effects of the non-axisymmetric potential. As some fundamental frequencies appearing in their solution are important for our discussions, we give here a very short description of their model.
They assume that the orbits of Pluto and Charon are Keplerian and circular around their center of mass. The binary is thus characterized by the mass of each body, respectively $m_p$ and $m_c$ ; by its separation $a_{\mathrm{bin}}$ and its mean motion $n_{\mathrm{bin}} = \sqrt{G \, \frac{m_p + m_c}{a_{\mathrm{bin}}^{3}}} $.
Using a cylindrical coordinate system ($R$, $\phi$, $z$) centered on the binary barycenter with the $z = 0$ plane as the orbital plane of the binary, the positions of Pluto and Charon are
\begin{equation}
    \vec{r}_p = \mqty(a_p \\\phi_c + \pi\\0) 
    \qquad 
    \vec{r}_c = \mqty(a_c \\\phi_c\\0)
\end{equation}
where $a_p = a_{\mathrm{bin}} \, \frac{m_c}{m_p + m_c}$, $a_c = a_{\mathrm{bin}} \, \frac{m_p}{m_p + m_c}$ and $\phi_c = n_{\mathrm{bin}} \, t + \varphi_{\mathrm{bin}}$.
\\
\\
One can then derive the time-depending, non-axisymmetric potential of the binary and put it into the equations of motion in cylindrical coordinates. So, the first-order approximation gives the following solution:
\begin{equation}
    \begin{aligned}
        \nu_k &= \sum_{i=1}^{15} \alpha_{k,i} \, \omega_i, \text{où} \, \alpha_{k,i} \subset \mathbb{Z} \\
        r_\text{s}(t) &= r_0 \left[ 1 - e \, cos\left(\kappa_\text{s} \, t + \psi \right) \sum_{k=1}^{\infty} C_k \, cos \left( k \left(n_\text{s} - n_\text{bin}\right) t + \Delta \phi\right)\right] \\
        R_\mathrm{s}(t) &= R_{\mathrm{s}, 0} \left[ 1 - e_s \cos(\kappa_\mathrm{s} t + \psi_\mathrm{s}) \displaystyle\sum_{k=1}^{\infty} C_{\mathrm{s}, k} \cos(k \Delta \theta_\mathrm{s}) \right]\\
        \phi_\mathrm{s}(t) &= n_\mathrm{s} \left[ t + 2 \frac{e_s}{\kappa_\mathrm{s}} \sin(\kappa_\mathrm{s} t + \psi_\mathrm{s}) \right. + \left. \displaystyle\sum_{k=1}^{\infty} \frac{D_{\mathrm{s}, k}}{k n_\mathrm{bin} - n_\mathrm{s}} \sin(k \Delta \theta_\mathrm{s})\right] + \varphi_\mathrm{s} \\
        z_\mathrm{s}(t) &= i_\mathrm{s} R_{\mathrm{s},0} \cos(\nu_\mathrm{s} t + \zeta_\mathrm{s})
    \end{aligned}
\label{for:cylindrical}
\end{equation}
avec $\Delta\theta_\mathrm{s} = (n_\mathrm{s} - n_\mathrm{bin})\,t + \varphi_\mathrm{s} - \varphi_\mathrm{bin}$. 
\\
\\
In the equations, the subscript $s$ denotes terms specific to either of the satellites in a circumbinary orbit (thus $s \in \{ \mathrm{Styx}, \mathrm{Nix}, \mathrm{Kerberos}, \mathrm{Hydra}\}$).
\\
\\
The motion in $R_s$ and $\phi_s$ is the superposition of the circular motion of a guiding center at $R_{s,0}$ with mean motion $n_s$, the epicyclic motion with amplitude $e_s$ and epicyclic frequency $\kappa_s$, and forced oscillations at multiples of the synodic frequency $\left| n_s - n_\mathrm{bin} \right|$. The motion in $z_s$ is decoupled from the previous one, with an amplitude $i_\mathrm{s} R_{\mathrm{s},0}$ and a vertical frequency $\nu_s$. The remaining parameters are integration constants.
Each of the three frequencies mentioned can be derived as truncated series based on the moon's distance from the binary's center of mass, as detailed in the Appendix of \citet{Bromley2021}.

\subsection{Numerical integrations} 
\label{integration_section}
Beyond the analytic model, we conduct simulations of the system to encompass greater complexity. Initially, we integrate a 3-body system consisting of the central binary and a single moon. The analytic model does indeed provide an accurate representation of this configuration. Subsequently, we extend our simulations to a complete 6-body system, allowing for the mutual interactions between the moons to be fully observed.
\\
\\
Since we do not consider the secular evolution of the system over timescales greater than a few Pluto's orbital period around the Sun, we use the IAS15 integrator within the N-body orbital integrator, REBOUND \citep{Rein2012, Rein2015}. It is a fast, adaptative, 15th order integrator, accurate to machine precision for billions of orbits, which exceeds by far our study's scope.
\\
\\
The initial conditions we use are the masses and state vectors of the latest solution provided by the JPL NAIF group, labeled PLU058\footnote{\url{https://naif.jpl.nasa.gov/pub/naif/generic_kernels/spk/satellites/} ; file: plu058.bsp}. This solution extends the one computed by \citet{Brozovic2015}, fitting the astrometry data available until 2021. However, it is only a best-fit solution and there is no uncertainty analysis provided. The uncertainties are particularly large for the masses of the two smallest moons, Styx and Kerberos, due to limited data and their marginal influence on the system. We detail the state vectors used in Table \ref{table:state_vectors}. The epoch is January 1, 2015 barycentric dynamic time (TDB).

\begin{table}[ht]
\centering
\caption{State vectors of Pluto's system.}
\begin{tabular}{lrr}
\hline \addlinespace[1pt]
Object & \multicolumn{1}{c}{Position (km)} & \multicolumn{1}{c}{Velocity (km s$^{-1}$)}\\ \addlinespace[1pt]
\hline \addlinespace[2pt]
Pluto & -1\,468.136\,624\,111\,64 & 0.006\,237\,219\,451\,353 \\
 & -1\,228.879\,498\,037\,82 & 0.009\,060\,674\,242\,529 \\
 & 937.812\,780\,490\,19 & 0.021\,637\,209\,578\,722 \\[5pt]
Charon & 12\,030.182\,705\,637\,53 & -0.051\,105\,142\,092\,561 \\
 & 10\,069.775\,067\,031\,02 & -0.074\,239\,958\,547\,253 \\
 & -7\,683.934\,690\,398\,31 & -0.177\,290\,246\,943\,712 \\[5pt]
Styx & -2\,273.632\,688\,746\,98 & 0.112\,669\,436\,117\,034 \\
 & 4\,171.376\,291\,830\,87 & 0.104\,445\,331\,033\,535 \\
 & 42\,110.759\,647\,305\,97 & -0.003\,603\,620\,202\,871 \\[5pt]
Nix & 18\,944.078\,817\,784\,89 & -0.087\,857\,174\,097\,601 \\
 & 11\,206.680\,100\,916\,48 & -0.091\,532\,691\,434\,227 \\
 & -43\,587.326\,446\,157\,09 & -0.063\,825\,714\,718\,057 \\[5pt]
Kerberos & -40\,918.972\,987\,952\,92 & 0.022\,200\,926\,576\,981 \\
 & -35\,829.432\,488\,240\,26 & 0.039\,703\,721\,742\,599 \\
 & 19\,157.062\,569\,099\,19 & 0.122\,450\,625\,504\,426 \\[5pt]
Hydra & -45\,810.720\,305\,207\,50 & 0.022\,653\,935\,330\,384 \\
 & -39\,726.392\,530\,275\,17 & 0.038\,638\,811\,170\,224 \\
 & 22\,811.806\,132\,549\,61 & 0.114\,570\,576\,175\,266 \\[5pt]
\hline
\end{tabular}
\floatfoot{Positions ($x$,$y$,$z$) and velocity ($v_x$,$v_y$,$v_z$) are given in the International
Celestial. Reference Frame (ICRF) cartesian coordinates relative to the Pluto system
barycenter. The epoch for the state vectors is January 1, 2015 TDB.}
\label{table:state_vectors}
\end{table}

\subsection{Frequency analysis}

\subsubsection{Integrable system and proper frequencies}

We know that a conservative dynamical system can be described by its frequencies \citep{Laskar1992}. We consider an integrable Hamiltonian system $H$ with $n$ degrees of freedom. If the system evolves within the hypothesis of the Arnold-Liouville theorem, some coordinates called action-angle $(J,\theta) $ exist for which the evolution is quite simple:       
\begin{equation}  
        H(\boldsymbol{J},{\boldsymbol{\theta}})=H_0 (J) \Longrightarrow  \begin{cases} \boldsymbol{J}(t)=\boldsymbol{J}_0 \\ {\boldsymbol{\theta}}(t)={\boldsymbol{\omega_J}} t+{\boldsymbol{\theta}}_0  \end{cases} \  
        (\boldsymbol{J},{\boldsymbol{\theta}})\in \mathbb{R}^n \times \  \mathbb{T}^n.  
        \label{for:act_ang}
\end{equation}

We note $J_i$ and $\omega_i$ for $i=1, \ldots , n$ respectively for the components of $\boldsymbol{J}$ and $\boldsymbol{\omega_J}$. The dynamics can be described by the complex variables $J_i \exp \left(\sqrt{-1} \theta_i(t)\right)$. The motion takes place on a $n$ dimension torus, with the radius  $J_i$, with the constant angular velocities $\omega_i$. Unfortunately, in most cases we don't know the form of these variables. Sometimes analytical theories can provide some that are close to them. However, if the system is integrable, the action-angle coordinates are intrinsic of the system. It means that, no matter the coordinate system we use, these coordinates will evolve with the proper frequencies $\omega_i$.

\subsubsection{The Frequency Map Analysis (FMA)}
\label{section:FMA}
Given the significance of these fundamental frequencies, we aim to determine their numerical values using a method more accurate than FFT, thereby achieving several orders of magnitude improvement in accuracy. The FMA program was written in C language by \citet{Saillenfest2014}.
The method is based on the works of \citet{Laskar1992} and \citet{Laskar1993}.
\\
\\
We define a function $f(t)$ that represents one of the time-dependent coordinates that come from the numerical integration of Sect.~\ref{integration_section}.
Then it is sampled with a given step $\Delta t$, small enough to allow neglecting both the numerical errors in the following quadratures and also the aliasing. Any frequency larger than the cut frequency $\nu_{\mathrm{cut}} $ cannot be correctly determined:
        \begin{equation}  
        \nu_{\mathrm{cut}} = \frac{\pi}{\Delta t}
        \label{for:nu_cut}\end{equation}        
Furthermore, $f(t)$ is limited to the interval $ [0, T] $ which defines a resolution frequency:
        \begin{equation}  
        \nu_0= \frac{2 \pi}{T}
        \label{for:nu_0}\end{equation}
We now suppose that $f(t)$ has a quasi-periodic form: 
        \begin{equation}
        f(t)= \sum_{k=1}^{N} A_k\, \exp\left( \sqrt{-1} ( {\nu}_{k}t  + {\phi}_k )\right) \,
        \text{\, with $A_k$, $\nu_k$ and $\phi_k \in \mathbb{R}$ }
        \label{for:quasi_per}\end{equation}
Each $\nu_k$ is an integer combination of the proper frequencies  $\omega_i$.                
The numerical determination of the frequencies $\nu_k$ is equivalent to finding the maxima of the function
$ \lvert \mathbb{A}(\nu) \rvert $ where:
 \begin{equation}
\mathbb{A}(\nu) = {\frac{1}{T}} \int^T _0 f(t) \, \exp\,(- \sqrt{-1} \, \nu t)   \, \mathrm{d}t,
\label{for:amplitude}
\end{equation} 
Usually, a Hanning window is added in this integral (Saillenfest 2014). In practice, we first performed a FFT in order to approximatively locate the first maximum and then computed the integral (\ref{for:amplitude}) around. The other frequencies are determined in the same way.
The precision of the determination is estimated to:
        \begin{equation}  
        \nu_{\mathrm{det}} = 2 \nu_0 \sqrt{\epsilon}
        \label{for:nu_det}\end{equation}
 where $\epsilon$ is the machine precision (here $\sim 10^{-15}$). With a time period of $T = 10^5$ days, the precision achieved in locating a peak is $\nu_\mathrm{det} \approx 10^{-12}$. If the peak is well isolated, meaning it is sufficiently separated from other peaks with similar amplitude, then the uncertainty corresponds to this localization error. In this case, a FMA is significantly more precise than a FFT, where the uncertainty corresponds to the resolution frequency $\nu_0$ which is several orders of magnitude less precise than $\nu_\mathrm{det}$ (here, $\nu_0 = 2 \pi \times 10^{-5}$).

\subsubsection{Identification of the frequencies in the Pluto-Charon system}
In our 6-body system within a 3-dimensional space, there are 15 degrees of freedom once we account for the motion of the barycenter. Therefore, we have $n = 15$ independant frequencies $\omega_i$, or fundamental frequencies. One of these corresponds to the conservation of angular momentum, which is related to the system's invariant plane: this is the zero frequency, which we denote as *.
\\
Three frequencies are associated with each of the 4 satellites, all of which follow a circumbinary orbit. According to the analytical model by Lee and Peale, these frequencies are:
\begin{itemize}
    \item[$\cdot$] $n_s$, the mean motion
    \item[$\cdot$] $\kappa_s$, the epicycic frequency
    \item[$\cdot$] $\nu_s$, the vertical motion frequency
\end{itemize} where the subscript $s$ denotes each satellite, as previously mentioned.
\\
\\
Lastly, the remaining two frequencies correspond to the binary's mean motion $n_\mathrm{bin}$ and its orbit's precession $\dot{\varpi}^*_\mathrm{bin}$. However, this model does not account for any potential resonances within the system. If one of these 15 frequencies is identified as an integer linear combination of the others, it will need to be replaced to establish an independant set of fundamental frequencies.
\\
\\
We recall that these 15 frequencies $\omega_i$ (including the zero frequency) are intrinsic to our dynamic system, regardless of the coordinates used to describe it. Obviously, and it is essential to highlight, the $\omega_i$ come from a choice in the designation of the frequencies found in the solution: any set of independent integer combinations of $\omega_i$ is also suitable. Of course, some choices are more fitting than others to better understand the dynamics, such as their physical or geometric significance, comparison with an analytical representation, and so on.
\\
Typically, to describe an object's motion in orbit, we use orbital elements, which change slowly over time. However, in this case, the standard orbital elements oscillate rapidly, making them ill-suited to describe a circumbinary orbit.
\\
Since we have an analytical model, as described in Sect.~\ref{epicyclic_section}, we will use it to identify the frequencies associated with each satellite. These frequencies appear in the motion equations in cylindrical coordinates (Equation \ref{for:cylindrical}). To extract them, we perform a frequency analysis on the time series $r_s$, $r_s \, \mathrm{e}^{\mathrm{i} \theta}$ and $z_s$. We will start by estimating these frequencies using the epicyclic model of \citet{Lee2006}, then identify them in a 3-body simulation, and finally in a 6-body simulation.

\subsubsection{Uncertainties}

\begin{figure}[ht!]
    \begin{tabular}{@{}c@{}}
    \resizebox{0.55\hsize}{!}{\includegraphics{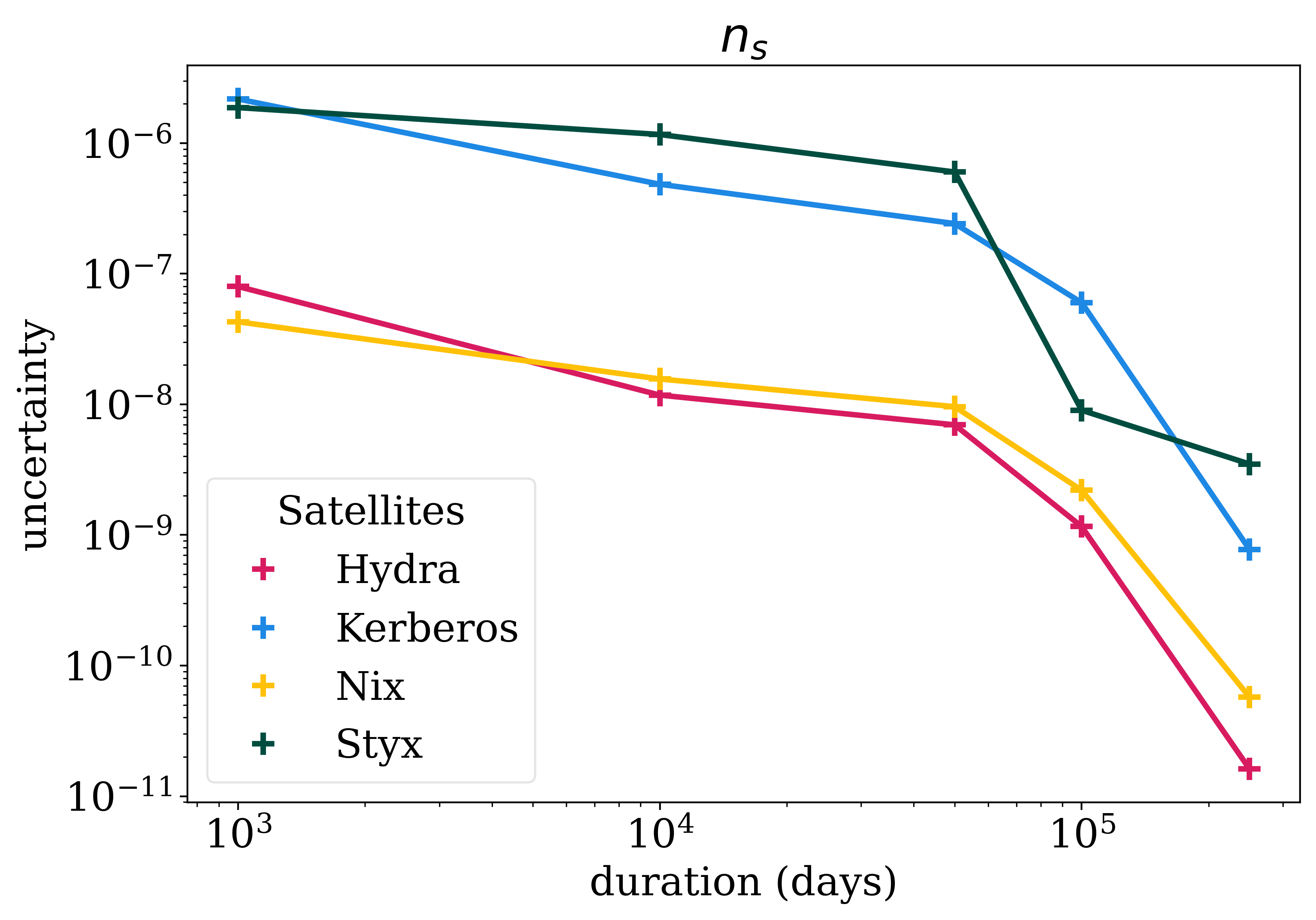}}
    \end{tabular}
    \begin{tabular}{@{}c@{}}
    \resizebox{0.55\hsize}{!}{\includegraphics{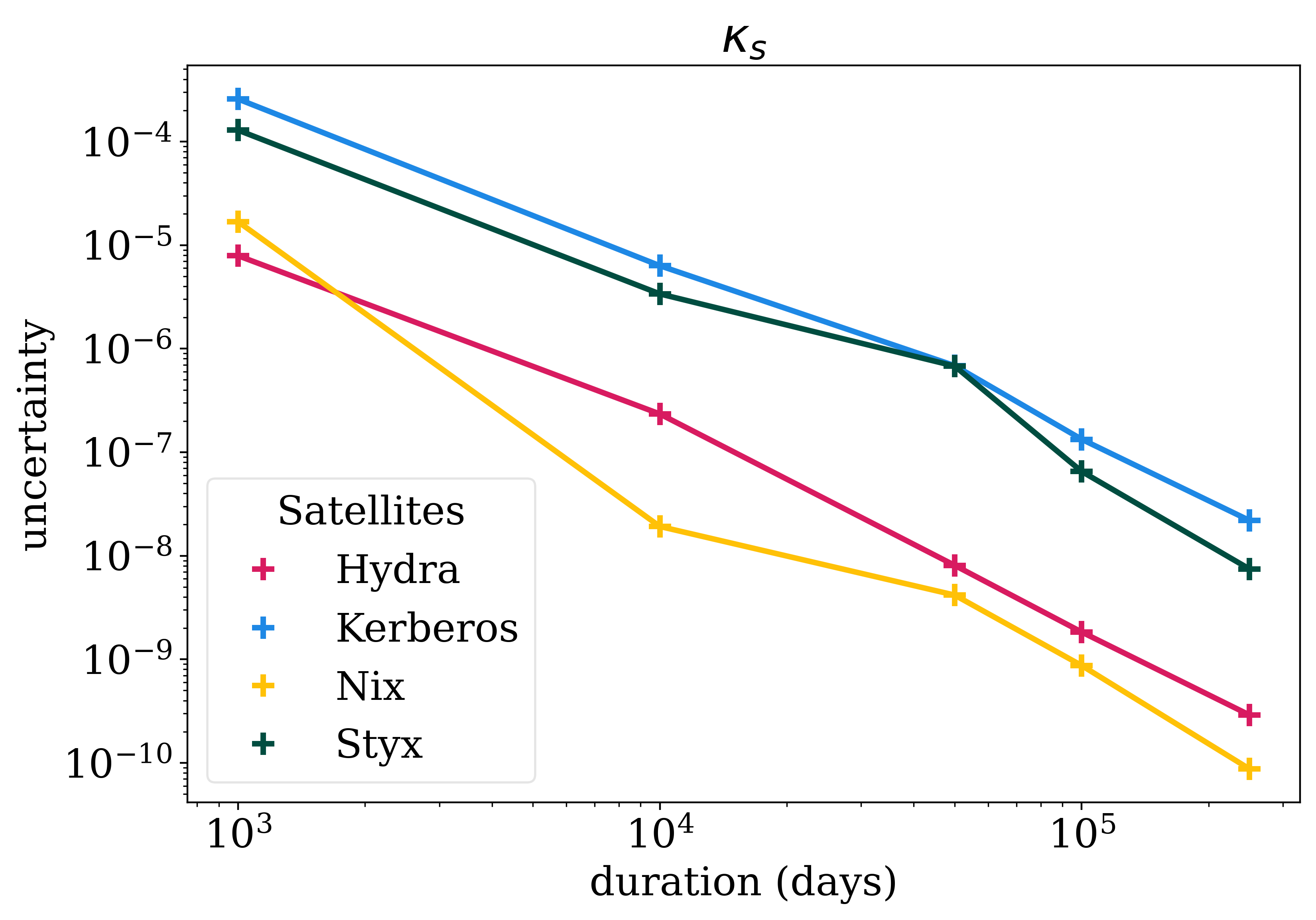}}
    \end{tabular}
    \begin{tabular}{@{}c@{}}
    \resizebox{0.55\hsize}{!}{\includegraphics{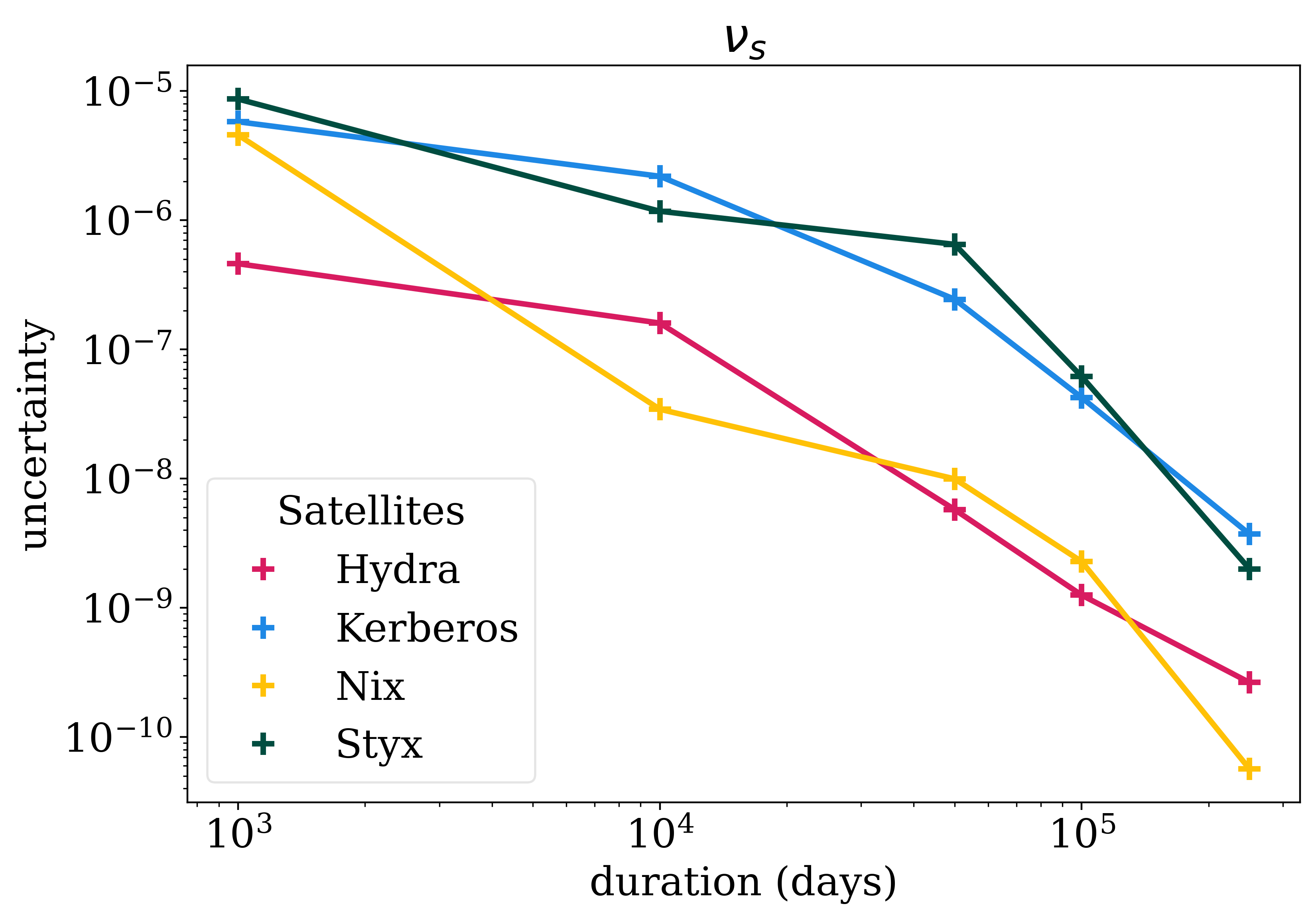}}
    \end{tabular}
\caption{Uncertainties associated to each fundamental frequency with the frequency analysis method. Mean motion frequency $n_s$ is associated to the dominant term in $r_s \, \mathrm{e}^{\mathrm{i} \theta}$. Epicyclic frequency $\kappa_s$ corresponds to the dominant term in $r_s$ for all satellites but Styx, which has other terms with a similar amplitude. Vertical motion frequency $\nu_s$ is associated to the dominant term in $z_s$. For each satellite and frequency, the analyses were performed with a duration of $1\,000$, $10\,000$, $50\,000$, $100\,000$ and $250\,000$ days.}
\label{ref:error_fig}
\end{figure}
The frequency analysis allows the identification of each term in the series decomposition of a quasi-periodic signal, in descending order of their amplitudes. Terms associated with well-isolated peaks and dominant amplitude are determined with higher precision due to minimal disturbance from other terms. The uncertainty associated with each frequency determined is thus inherently dependent on the temporal signal analyzed $T$.
\\
Moreover, certain long period effects such as nodal precession or periapse precession, as well as possible secular resonances, can occur over longer timescales. If these effects manifest themselves with peaks that are too close together, they can only be distinguished by reducing the resolution frequency $\nu_0$.
\\
\noindent
To determine the uncertainty associated with each term, a series of frequency analyses is performed on temporal signals of the same duration. Each of these signals is obtained by simulating the complete Pluto-Charon system, choosing evenly spaced initial dates. The value derived with the frequency analysis oscillates due to the poor determination of low-frequency effects. Therefore, the uncertainty associated with the determined frequencies corresponds to the amplitude of these oscillations.
\\
\\
Fig.~\ref{ref:error_fig} shows that, as expected, the uncertainty decreases as the duration of the analyzed signal increases. Regardless of the satellite considered, the mean motion $n_s$ gets the lowest uncertainties, followed by the vertical frequency $\nu_s$ and then the epicyclic frequency $\kappa_s$. Also, there is a significant perfomance gap between the smaller satellites Styx and Kerberos when compared to Nix and Hydra, with discrepancies spanning $1$ to $2$ orders of magnitude across all frequencies. Indeed, they are more likely to be affected by secular effects. 
\\
For both $\kappa_s$ and $\nu_s$, the log-linear plots show a predominantly linear trend with a large slope, although it is less pronounced for $\nu_s$. However, the uncertainties remain nearly constant up to $50\,000$ days for mean motions, after which there is a marked inflection point; after that, it decreases linearly. This suggests the potential influence of secular effects on these timescales.
\\
Lastly, the figure does not show the uncertainty for the frequencies associated to the binary itself. The frequency spectrum of Charon's variable $r \, \mathrm{e}^{\mathrm{i}\theta}$ is dominated by the binary's mean motion $n_\mathrm{bin}$, which always reach the minimum uncertainty $\nu_\mathrm{det}$ for any duration used for the analysis. The next significant frequency within the spectrum is associated with $\dot{\varpi}^*_\mathrm{bin}$. It is extremely low, way under the frequency resolution $\nu_0$ but still above $\nu_\mathrm{det}$. It corresponds to a period of tens of millions of days, which means that further investigation would require to take into account the effects of the Solar System, that operate on such a timescale.
\\
\\
Reproducing the whole analysis on different time series, such as non-singular Keplerian elements, yielded mostly the same results.

%
\section{The fundamental frequencies of the system} \label{freq_section}

 We now compare the values obtained throughout the different methods exposed in Sect.~\ref{integration_section}. The numerical integrations were all conducted with a steptime $\Delta t = 0.5$ days and a duration $T = 250\,000$ days.
 \\
 \\
 Table \ref{table:compare} shows the results yield by each method for all the 12 satellites' frequencies. For the sake of clarity, we used the following transformations: 
 \begin{itemize}
     \item[$\cdot$] $\dot{\varpi}^*_s = n_s - \kappa_s$, the apsidal precession rate.
     \item[$\cdot$] $\dot{\Omega}^*_s = n_s - \nu_s$, the nodal precession rate.
 \end{itemize}
 When $e_s \gg \sum_{k=1}^{\infty} C_k$, the orbit can effectively be described as a precessing Keplerian ellipse, with a periapse precession rate $\dot{\varpi}_s \approx \dot{\varpi}^*_s$. Additionnaly, if the inclination with respect to the Pluto-Charon orbital plane $i \neq 0$, the ellipse plane precesses with $\dot{\Omega}_s \approx \dot{\Omega}^*_s$.
\\
\\
\noindent
We observe a strong agreement between the frequencies derived from the analytical model and the numerical simulations, even in a 6-body scenario. 
\\
However, the discrepancy between the periapsis precession frequency $\dot{\varpi}^*_\mathrm{Styx}$ of the model and the one obtained from the frequency analysis becomes more pronounced, especially when the satellite is closer to the central binary. This was expected, as demonstrated by \citet{Woo2020}.

\begin{table}[]
\centering
\caption{Comparison of frequencies values from Epicyclic Theory and Frequency Analysis.}
\begin{tabular}{lrrr}
\hline \addlinespace[1pt]
Method & \multicolumn{1}{c}{$n_\mathrm{Styx}$} & \multicolumn{1}{c}{$\dot{\varpi}^*_\mathrm{Styx}$} & \multicolumn{1}{c}{$\dot{\Omega}^*_\mathrm{Styx}$} \\ \addlinespace[1pt]
\hline \addlinespace[2pt]
Epicyclic & 0.311\,744 & 0.006\,504 & -0.006\,371 \\
3-body    & 0.311\,625 & 0.002\,670 & -0.006\,537 \\
6-body    & 0.311\,617 & 0.002\,676 & -0.006\,541 \\[5pt]
\hline \addlinespace[1pt]
Method & \multicolumn{1}{c}{$n_\mathrm{Nix}$} & \multicolumn{1}{c}{$\dot{\varpi}^*_\mathrm{Nix}$} & \multicolumn{1}{c}{$\dot{\Omega}^*_\mathrm{Nix}$} \\ \addlinespace[1pt]
\hline \addlinespace[2pt]
Epicyclic & 0.252\,811 & 0.003\,717 & -0.003\,663 \\
3-body    & 0.252\,782 & 0.003\,307 & -0.003\,670 \\
6-body    & 0.252\,781 & 0.003\,308 & -0.003\,671 \\[5pt]
\hline \addlinespace[1pt]
Method & \multicolumn{1}{c}{$n_\mathrm{Kerberos}$} & \multicolumn{1}{c}{$\dot{\varpi}^*_\mathrm{Kerberos}$} & \multicolumn{1}{c}{$\dot{\Omega}^*_\mathrm{Kerberos}$} \\ \addlinespace[1pt]
\hline \addlinespace[2pt]
Epicyclic & 0.195\,323 & 0.001\,911 & -0.001\,892 \\
3-body    & 0.195\,313 & 0.001\,861 & -0.001\,894 \\
6-body    & 0.195\,312 & 0.001\,869 & -0.001\,901 \\[5pt]
\hline \addlinespace[1pt]
Method & \multicolumn{1}{c}{$n_\mathrm{Hydra}$} & \multicolumn{1}{c}{$\dot{\varpi}^*_\mathrm{Hydra}$} & \multicolumn{1}{c}{$\dot{\Omega}^*_\mathrm{Hydra}$} \\ \addlinespace[1pt]
\hline \addlinespace[2pt]
Epicyclic & 0.164\,463 & 0.001\,240 & -0.001\,231 \\
3-body    & 0.164\,461 & 0.001\,226 & -0.001\,231 \\
6-body    & 0.164\,462 & 0.001\,227 & -0.001\,232 \\[5pt]
\hline
\end{tabular}
\floatfoot{Epicyclic theory values are computed using the formulas provided within the Appendix of \citet{Bromley2021}. All values are given with 6 digits precision to better highlight the differences between each method. All values are given in $2\pi \cdot \text{days}^{-1}$}
\label{table:compare}
\end{table}

\begin{table}[]
\centering
\caption{Fundamental frequencies of the Pluto-Charon system}
\begin{tabular}{ll}
\hline \addlinespace[1pt]
\multicolumn{1}{c}{Frequency} & \multicolumn{1}{c}{Value ($2\pi \cdot \text{days}^{-1}$)}\\ 
\addlinespace[1pt] \hline \addlinespace[2pt]
$\text{*}$                      &  $\text{\,\,0}$ \\
\addlinespace[2pt] \hline \addlinespace[2pt]
$n_\mathrm{bin}$                  &  $\text{\,\,0.983\,646\,194\,025}$ \\
$\dot{\varpi}_\mathrm{bin}$       &  $\text{\,\,0.000\,000\,089\,675}$ \\
\addlinespace[2pt] \hline \addlinespace[2pt]
$n_\mathrm{Styx}$                 &  $\text{\,\,0.311\,616\,87\underline{9\,655}}$ \\
$\dot{\varpi}^*_\mathrm{Styx}$      &  $\text{\,\,0.002\,676\,27\underline{9\,375}}$ \\
$\dot{\Omega}^*_\mathrm{Styx}$      &  $\text{-0.006\,541\,04\underline{9\,126}}$ \\
\addlinespace[2pt] \hline \addlinespace[2pt]
$n_\mathrm{Nix}$                  &  $\text{\,\,0.252\,780\,973\,5\underline{63}}$ \\
$\dot{\varpi}^*_\mathrm{Nix}$       &  $\text{\,\,0.003\,307\,892\,1\underline{43}}$ \\
$\dot{\Omega}^*_\mathrm{Nix}$       &  $\text{-0.003\,670\,532\,8\underline{95}}$ \\
\addlinespace[2pt] \hline \addlinespace[2pt]
$n_\mathrm{Kerberos}$             &  $\text{\,\,0.195\,312\,103\,\underline{061}}$ \\
$\dot{\varpi}^*_\mathrm{Kerberos}$  &  $\text{\,\,0.001\,868\,66\underline{2\,023}}$ \\
$\dot{\Omega}^*_\mathrm{Kerberos}$  &  $\text{-0.001\,901\,180\,\underline{669}}$ \\
\addlinespace[2pt] \hline \addlinespace[2pt]
$n_\mathrm{Hydra}$                &  $\text{\,\,0.164\,462\,100\,7\underline{57}}$ \\
$\dot{\varpi}^*_\mathrm{Hydra}$     &  $\text{\,\,0.001\,226\,841\,\underline{678}}$ \\
$\dot{\Omega}^*_\mathrm{Hydra}$     &  $\text{-0.001\,232\,211\,\underline{931}}$ \\
\addlinespace[2pt] \hline \hline \addlinespace[2pt]
$\nu_\mathrm{cut}$                &   $\text{\,\,1}$ \\
$\nu_0$                         &   $\text{\,\,0.000\,004}$\\
$\nu_\mathrm{det}$                &   $\text{\,\,0.000\,000\,000\,000\,3}$ \\
\hline
\end{tabular}
\floatfoot{The values which characterize the frequency analysis ($\nu_\mathrm{cut}$, $\nu_0$ and $\nu_\mathrm{det}$) are recalled at the end of the table. The uncertainty associated to each frequency is shown with underlined digits.}
\label{table:fundamentals}
\end{table}

Table \ref{table:fundamentals} presents the complete set of 15 fundamental frequencies obtained with the 6-body simulation. All values are displayed with 12 digits as it corresponds to the maximum accuracy given by $\nu_\mathrm{det}$, for a duration $T = 250\,000$ days. We recall at the end of the table the values $\nu_\mathrm{cut}$, $\nu_0$ and $\nu_\mathrm{det}$, which characterize the frequency analysis as discussed in Sect.~\ref{section:FMA}.

%
\section{Commensurabilities in the Pluto-Charon system}
\label{section4}

Building on the frequencies computed in Sect.~\ref{freq_section}, we investigated the dynamic resonances within the Pluto-Charon system. These resonances are characterized by a specific relationship, denoted by the resonant angle $\Phi$ defined as:
\begin{equation}
    \Phi = \sum_s a_s \lambda_s + b_s \varpi_s + c_s \Omega_s
\label{equation:resonance}
\end{equation}
where $a_s$, $b_s$, $c_s$ are integer coefficients satisfying the condition $\sum_s \left(a_s + b_s + c_s\right) = 0$.
\\
\\
An active resonance is indicated by $\dot{\Phi} \approx 0$, with the resonant angle $\Phi$ usually librating around a given value, often $0^\circ$ or $180^\circ$.
\\
More precisely, we used the fundamental frequencies of Sect.~\ref{freq_section} to identify the integer coefficients $a_s$, $b_s$ and $c_s$ which satisfy that condition:
\begin{equation}
    \dot{\Phi}^* = \sum_s a_s \, n_s + b_s \, \dot{\varpi}^*_s + c_s \, \dot{\Omega}^*_s \approx 0
\label{equation:resonance_deriv}
\end{equation}
In contrast, when the resonance is inactive, the angle circulates.
\\
\\
Earlier work by \citet{Showalter2015} identified a potential 
three-body resonance involving Styx, Nix, and Hydra, characterized by the relation $\Phi = 3 \lambda_{\mathrm{Styx}} - 5 \lambda_{\mathrm{Nix}} + 2 \lambda_{\mathrm{Hydra}}$. Through numerical integrations, they suggested the masses of Nix and Hydra were underestimated so that the angle would librate. However, more recent estimations by \citet{Porter2023} argue that these masses were actually overestimated. 
Another three-body resonance was mentionned, with the relation $\Phi = 42 \lambda_{\mathrm{Styx}} - 85 \lambda_{\mathrm{Nix}} + 43 \lambda_\mathrm{Kerberos}$.

\subsection{Identifying candidates}
Our approach deviates from that of \citet{Showalter2015} by focusing on the semi-major axis of Styx to explore potential near-resonances. Styx is indeed the smallest satellite of the system and has the least constrained orbit. Slight adjustements in its semi-major axis won't significantly impact most frequencies but its own mean motion $n_\mathrm{Styx}$.
\\
We thus bound $n_\mathrm{Styx}$ between an upper and a lower value, based on the uncertainty in the semi-major axis $\Delta a$. For any linear combination with up to four frequencies, we retain those where a change of sign occurs, which signifies the transition of $\dot{\Phi}^*$ through zero. This approach allows us to identify combinations which are candidates for a long-period resonance.
\\
\\
Using the initial state vectors and masses, we compute the corresponding angle $\Phi$. An angle nearing $0^\circ$ or $180^\circ$ suggests a higher likelihood of an active resonance.
\\
\\
However, due to the circumbinary nature of the satellites' orbits, the longitude of the pericenter $\varpi^*_s$ is particularly ill-suited, as it exhibits erratic changes and circulates over a period commensurable with the mean longitude $\lambda_s$. 
\\
Therefore, we chose to focus on examples where the frequency $\dot{\varpi}^*_s$ does not appear.

\subsection{Results}

Using this method, we identified over 100 candidates for potential dynamic resonances within the Pluto-Charon system. Each candidate needs to be evaluated to determine whether the angle $\Phi$ is close
to libration.
\\
Our analysis prioritized combinations that most closely resemble the current system dynamics and feature the smallest integer coefficients. This approach helps us focus on the most probable resonances.
\\
\\
The main challenge in determining the behavior of $\Phi$ is to interprate its motion when its rate of change, $\dot{\Phi}^*$, is exceedingly small. In such cases, what might appear as linear, non-oscillatory motion over a given time span could, over an extended period, reveal itself to be oscillatory. Our capacity to simulate the system over long durations is limited, both for computational resources and to stay within the scope of the study. For instance, even a small perturbation from the Sun or the outer planets could cancel out long-term effects induced by a resonance.
\\
\\
In light of these considerations, we have selected a few candidates that showed promising outcomes, as presented in Table \ref{table:candidates}.

\begin{figure}[ht!]
    \begin{tabular}{@{}c@{}}
    \resizebox{0.55\hsize}{!}{\includegraphics{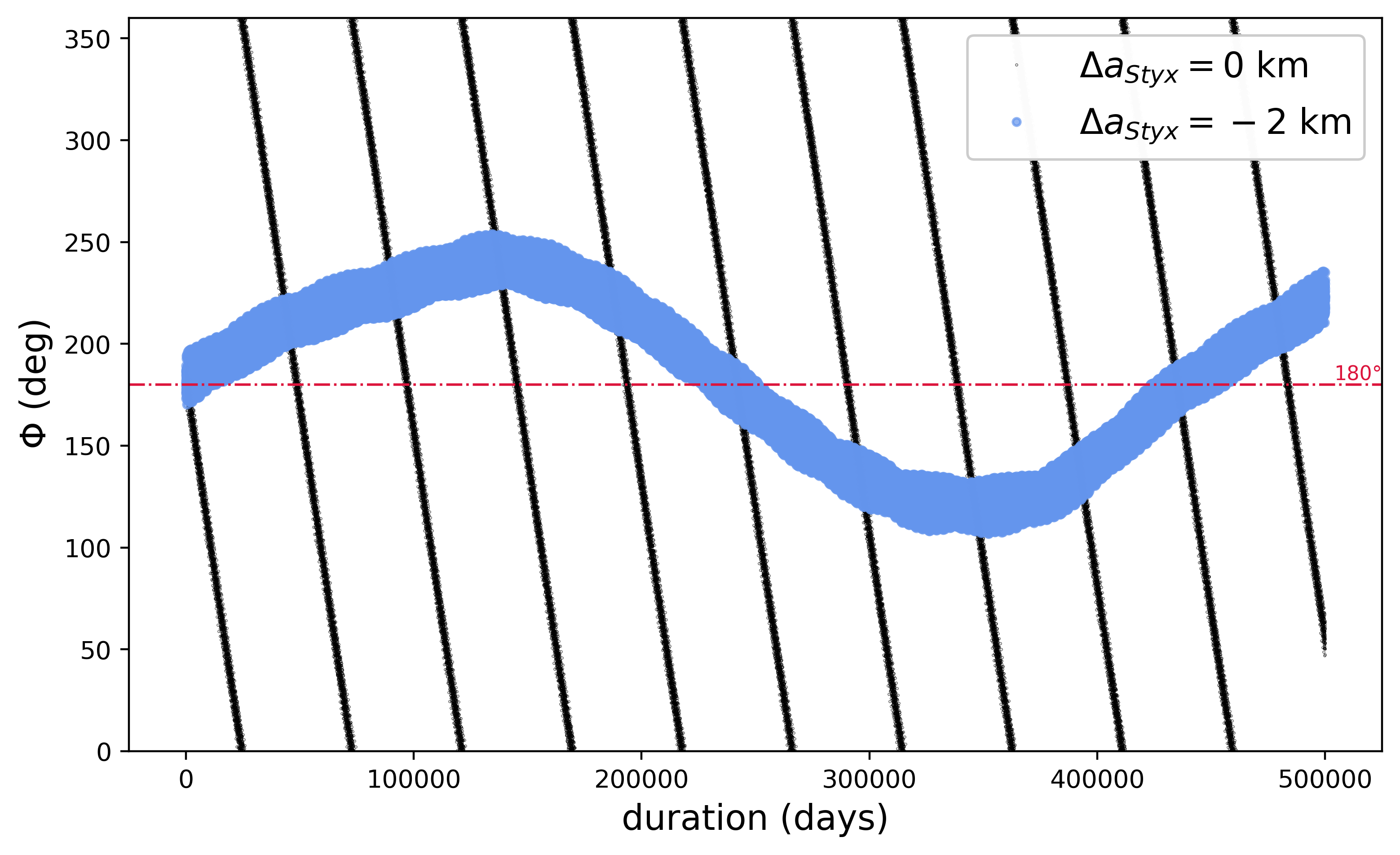}}
    \end{tabular}
\caption{Angle $\Phi$ evolution at commensurability $\Phi = 3 \lambda_\mathrm{Styx} - 5 \lambda_\mathrm{Nix} + 2 \lambda_\mathrm{Hydra}$.}
\label{fig:resonance3}
\end{figure}

\begin{table}[ht!]
\centering
\caption{More convincing potential near-resonances candidates}
\begin{tabular}{l}
\hline \addlinespace[1pt]
\multicolumn{1}{c}{Combination} \\ \addlinespace[1pt]
\hline \addlinespace[2pt]
$3 \lambda_\mathrm{Styx} - 5 \lambda_\mathrm{Nix} + 2 \lambda_\mathrm{Hydra}$ \\[2pt]
$2 \lambda_\mathrm{bin} - 5 \lambda_\mathrm{Nix} + 14 \lambda_\mathrm{Kerberos} - 11 \lambda_\mathrm{Styx}$ \\[2pt]
$2 \lambda_\mathrm{Styx} - 4 \lambda_\mathrm{Kerberos} + \lambda_\mathrm{Hydra} + \Omega_\mathrm{Styx}$\\[2pt]
$\lambda_\mathrm{bin} + 4 \lambda_\mathrm{Nix} + 3 \lambda_\mathrm{Hydra} - 8 \lambda_\mathrm{Styx}$ \\[2pt]
\hline
\end{tabular}
\label{table:candidates}
\end{table}
\noindent
In Figs. 2-5, the resonant angle $\Phi(t)$ associated to each of these candidates is presented for two different systems: one with $\Delta a = 0$, that is, with the initial conditions of Table \ref{table:state_vectors} and the second one with the indicated change in Styx's semi-major axis $\Delta a$.
\\
\\
The main three-body resonance found by \citet{Showalter2015}, $\Phi = 3 \lambda_\mathrm{Styx} - 5 \lambda_\mathrm{Nix} + 2 \lambda_\mathrm{Hydra}$, is shown in Fig.~\ref{fig:resonance3}. A slight change of $\approx$ 1-3 km in Styx's semi-major axis is enough to modify the dynamics of the angle $\Phi$. This indicates that the transition from circulation to libration spans a broad region.

\begin{figure}[t!]
    \begin{tabular}{@{}c@{}}
    \resizebox{0.55\hsize}{!}
    {\includegraphics{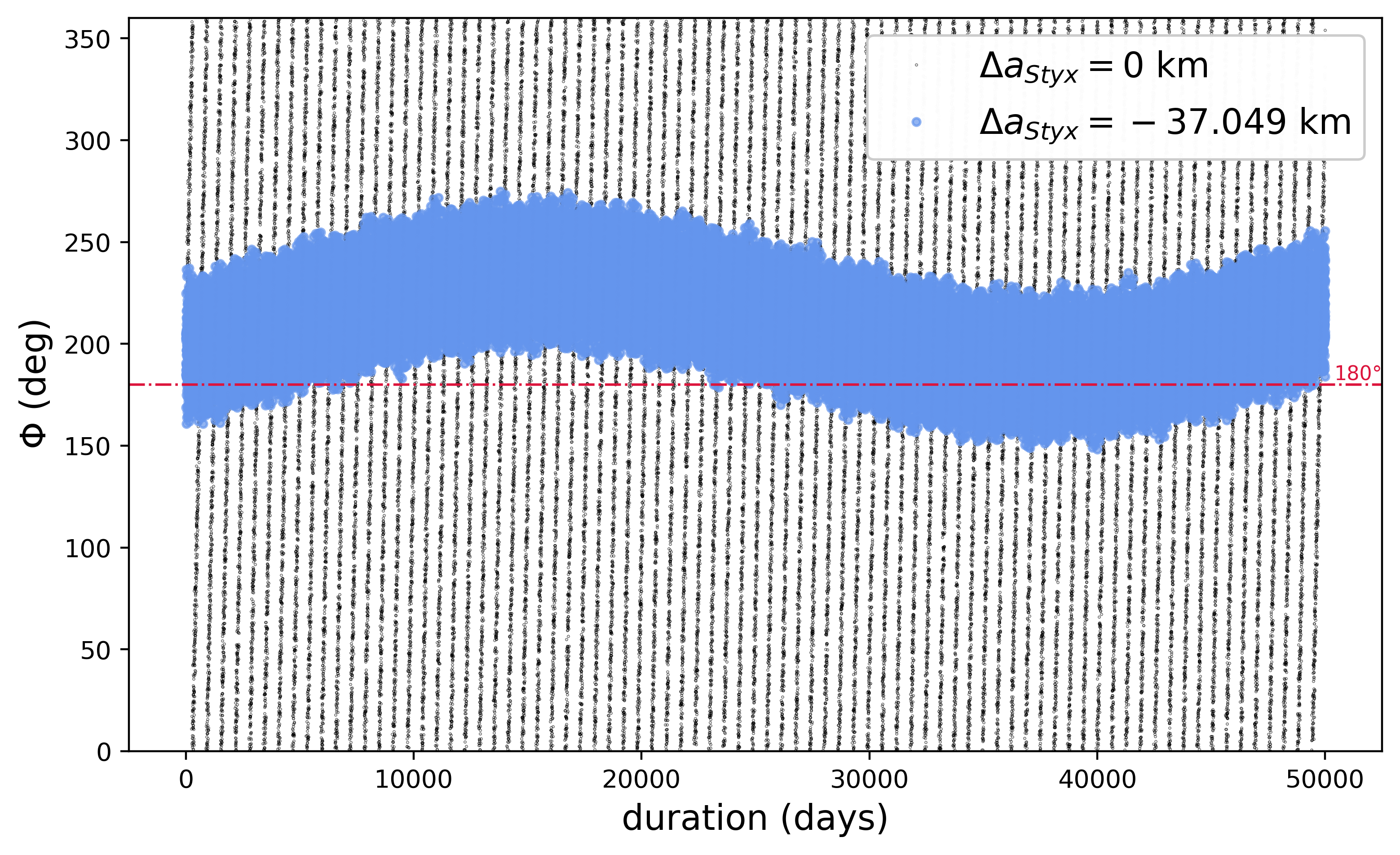}}
    \end{tabular}
\caption{Angle $\Phi$ evolution at commensurability $\Phi = 2 \lambda_\mathrm{bin} - 5 \lambda_\mathrm{Nix} + 14 \lambda_\mathrm{Kerberos} - 11 \lambda_\mathrm{Styx}$.}
\label{fig:resonance1}
\end{figure}

\begin{figure}[ht!]
    \begin{tabular}{@{}c@{}}
    \resizebox{0.55\hsize}{!}{\includegraphics{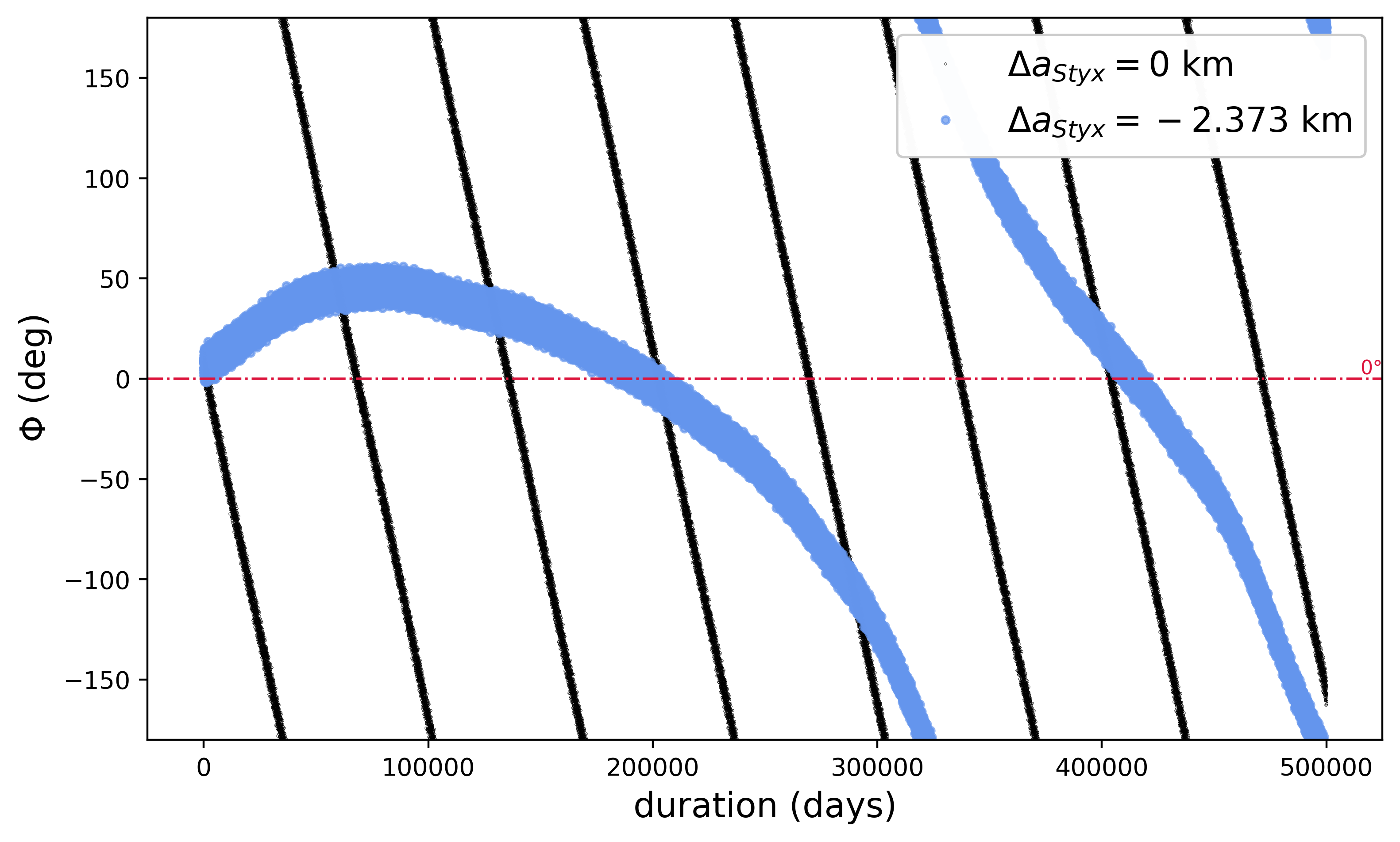}}
    \end{tabular}
\caption{Angle $\Phi$ evolution at commensurability $\Phi =  2 \lambda_\mathrm{Styx} - 4 \lambda_\mathrm{Kerberos} + \lambda_\mathrm{Hydra} + \Omega_\mathrm{Styx}$}
\label{fig:resonance4}
\end{figure}

\begin{figure}[t!]
    \begin{tabular}{@{}c@{}}
    \resizebox{0.55\hsize}{!}{\includegraphics{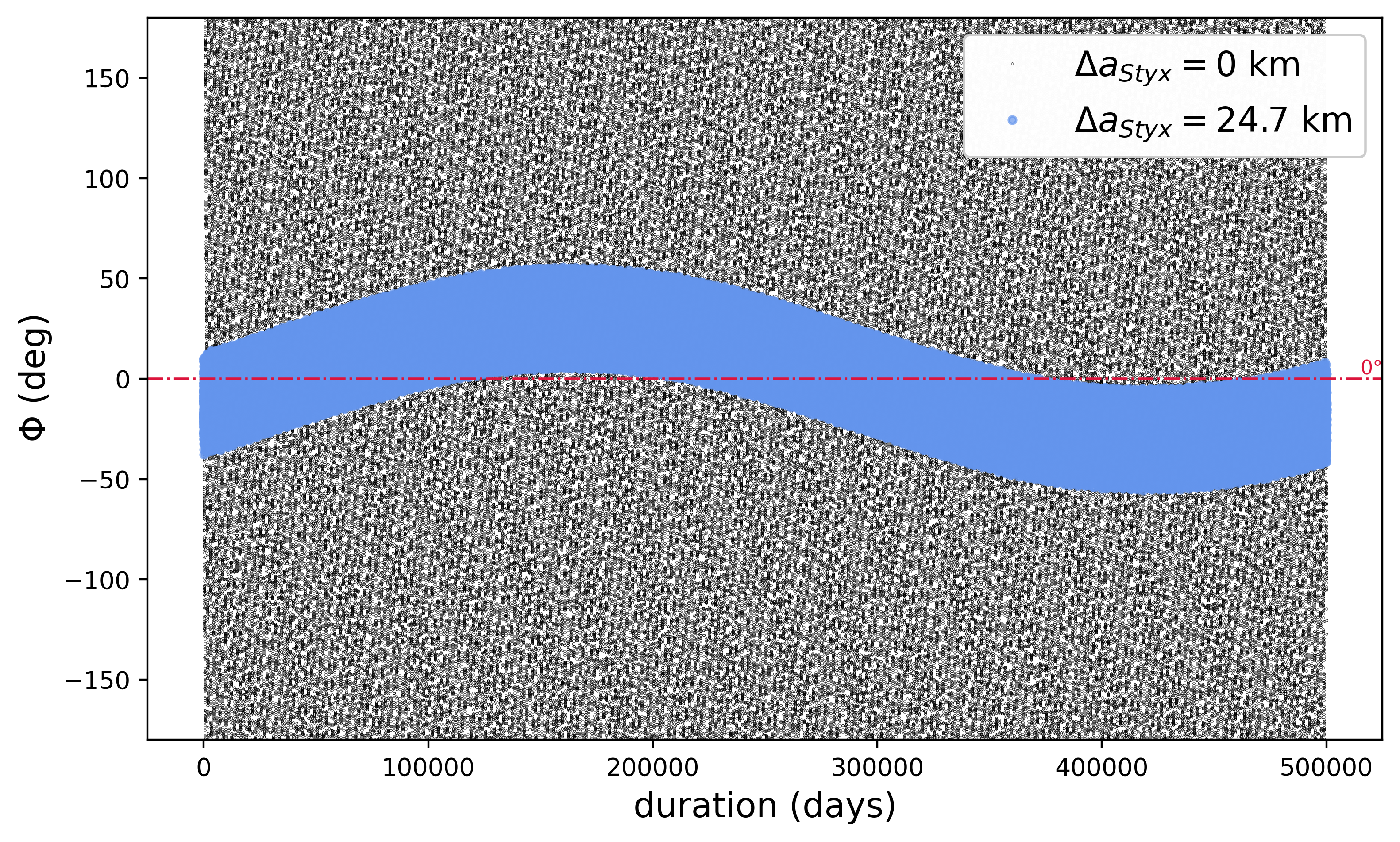}}
    \end{tabular}
\caption{Angle $\Phi$ evolution at commensurability $\Phi = \lambda_\mathrm{bin} + 4 \lambda_\mathrm{Nix} + 3 \lambda_\mathrm{Hydra} - 8 \lambda_\mathrm{Styx}$}
\label{fig:resonance2}
\end{figure}

\noindent
This is not the case for $\Phi = 2 \lambda_\mathrm{bin} - 5 \lambda_\mathrm{Nix} + 14 \lambda_\mathrm{Kerberos} - 11 \lambda_\mathrm{Styx}$, the mean motion resonance shown in Fig.~\ref{fig:resonance1}. Indeed, the transition between circulation and libration occurs within a very narrow region, located $\approx 37.049$ km away.
\\
\\
The resonance $\Phi = 2 \lambda_\mathrm{Styx} - 4 \lambda_\mathrm{Kerberos} + \lambda_\mathrm{Hydra} + \Omega_\mathrm{Styx}$ in Fig.~\ref{fig:resonance4} illustrates well that the resonant angle can have a more complex behavior between circulation and libration, which is typical for near-resonances. Once again, the system is close to the current value of Styx's semi-major axis.
\\
\\
Finally, Fig.~\ref{fig:resonance2} shows another near mean motion resonance case with the combination $\Phi = \lambda_\mathrm{bin} + 4 \lambda_\mathrm{Nix} + 3 \lambda_\mathrm{Hydra} - 8 \lambda_\mathrm{Styx}$, much narrower than the first example.

\noindent
Those examples highlight the system's sensitivity to minor orbital adjustments and suggest the existence of resonances that are dynamically significant, although they might be too narrow. Regarding the investigation of the second mean motion resonance mentioned earlier ($\Phi = 42 \lambda_{\mathrm{Styx}} - 85 \lambda_{\mathrm{Nix}} + 43 \lambda_\mathrm{Kerberos}$), we observed that despite $\dot{\Phi}^*$ being close to zero, there was no evidence of libration. Furthermore, this commensurability is not isolated; we identified other similar near-resonance candidates with nearly identical coefficients, such as $\Phi = 41 \lambda_{\mathrm{Styx}} - 83 \lambda_{\mathrm{Nix}} + 42 \lambda_\mathrm{Kerberos}$, among others. Such resonances are unlikely, due to the high coefficients involved in the combinations.

%

\section{Conclusion}

In this work we have studied the fundamental frequencies of the four satellites of the Pluto-Charon's system. We have used the epicyclic theory of \citet{Lee2006} and numerical integrations of the 3-body problem to devise our fine frequency analysis. Then, numerical integrations of the 6-body problem allows us to obtain all the 15 fundamental frequencies of the system, with a far greater precision than obtained before with a FFT technique. Our method can be used even with a more valuable model.
\\
\\
The mass modifications proposed by \citet{Showalter2015}, aimed at activating a resonance, is not consistent with the more accurate mass determinations provided by \citet{Porter2023}. Instead, we introduced another approach: by slightly altering the initial conditions, it is easy to observe the libration of the argument corresponding to the commensurability. In the examples presented herein, a minor change to the semi-major axis of Styx suffice to activate the resonances. However, our analysis does not quantify the precise extent of these resonant zones. We confirmed that the system is in the vicinity of a three-body resonance involving Styx, Nix and Hydra, as identified by \citet{Showalter2015}, characterized by the relation $\Phi = 3 \lambda_\mathrm{Styx} - 5 \lambda_\mathrm{Nix} + 2 \lambda_\mathrm{Hydra}$.
\\
\\
The assumptions made for the simulations performed herein might have introduce some discrepancies between our results and the real dynamics of the Pluto-Charon system. One way to tackle this problem would be to account for more perturbations, such as the tidal effects. The long-term dynamics would also benefit from a greater timespan simulation. The use of real observational data, although limited in time, could help find if the system is near resonances or at the exact location it must be to activate some of them.
The effects these resonances have on the system dynamics are yet to be explored.

\bibliographystyle{unsrtnat}
\bibliography{references}
\end{document}